\def\Tr{{\text{Tr}}\,}

\def\be{\begin{equation}}
\def\ee{\end{equation}}
\def\bea{\begin{eqnarray}}
\def\eea{\end{eqnarray}}
\def\bse{\begin{subequations}}
\def\ese{\end{subequations}}

\documentclass[prl,twocolumn,showpacs,preprintnumbers,amsmath,amssymb]{revtex4}
\usepackage{graphicx}
\usepackage{dcolumn}
\usepackage{bm}
\begin{document}
\title{Breakdown of Hydrodynamic Transport Theory in the Ordered Phase of
       Helimagnets
}
\author{T.R. Kirkpatrick$^{1}$ and D. Belitz$^{2}$}
\affiliation{$^{1}$Institute for Physical Science and Technology and Department
                   of Physics, University of Maryland, College Park, MD 20742\\
             $^{2}$Department of Physics, Institute of Theoretical Science, and
                   Materials Science Institute, University of Oregon, Eugene, OR
                   97403}
\date{\today}
\begin{abstract}
It is shown that strong fluctuations preclude a hydrodynamic description of
transport phenomena in helimagnets, such as MnSi, at $T>0$. This breakdown of
hydrodynamics is analogous to the one in chiral liquid crystals. Mode-mode
coupling effects lead to infinite renormalizations of various transport
coefficients, and the actual macroscopic description is nonlocal. At $T=0$
these effects are weakened due to the fluctuation-dissipation theorem, and the
renormalizations remain finite. Observable consequences of these results, as
manifested in the neutron scattering cross-section, are discussed.
%
\end{abstract}


\maketitle

Soft modes, or excitations at low frequencies and small wave numbers, in
many-body systems are crucial for the behavior at long distances and times.
They also lead to characteristic nonanalyticities in the temperature ($T$)
dependence of observables that are given by integrals over the excitation
spectrum, such as the specific heat. Examples include shear diffusion in
classical fluids \cite{Boon_Yip_1991}, the `diffuson' in quenched disordered
electron systems at $T=0$ \cite{Lee_Ramakrishnan_1985}, magnons in magnetically
ordered solids \cite{Kittel_1996}, the director fluctuations in liquid crystals
\cite{deGennes_Prost_1993}, and many others. Interactions between such soft
modes, often referred to as mode-mode coupling effects, lead to even more
interesting phenomena. In classical fluids, they produce the so-called
long-time tails, or non-exponential decay of time correlation functions, e.g.,
the transverse velocity autocorrelation function
\cite{Kirkpatrick_Belitz_Sengers_2002}. In smectic and cholesteric liquid
crystals they are even stronger, and lead to a breakdown of hydrodynamics, with
certain viscosities diverging for vanishing frequency $\omega$ as $1/\omega$
\cite{Mazenko_Ramaswamy_Toner_1983}. The weak-localization anomalies in
disordered electron systems \cite{Lee_Ramakrishnan_1985} belong to the same
category of effects, although they are usually not cast in this language.
Technically, the mode-mode coupling effects take the form of nonlinearities in
dynamical equations, or loops in a field-theoretic description.

The above examples make it clear that mode-mode coupling effects, and the
singularities induced by them, are relevant for both classical and quantum
many-body systems. An obvious question is therefore what happens to the
classical singularities as $T\to 0$. From the structure of the classical
singularities it is likely that they become weaker with decreasing temperature,
while quantum singularities may develop that become stronger, but this has
never been investigated.

One purpose of the present Letter is to study an explicit example of what
happens to classical mode-mode coupling singularities as $T\to 0$. Our example
are helimagnets, such as MnSi, which have recently been shown to exhibit
interesting analogies with smectic and cholesteric liquid crystals. These
analogies exist because the order parameters in the two systems are essentially
identical and the symmetries of the respective coarse-grained free-energy
functionals are closely related. Consequently, a tree-level analysis of both a
phenomenological and a microscopic theory of helimagnets
\cite{Belitz_Kirkpatrick_Rosch_2006} reveals a Goldstone mode, and its
manifestation in observables, that are very similar to the corresponding
properties of chiral liquid crystals \cite{deGennes_Prost_1993, Lubensky_1972}.
These properties are very unusual because of the anisotropic nature of the
helical Goldstone mode, which is softer in the direction transverse to the
pitch vector of the helix than in the longitudinal direction.

A second purpose of this Letter is to investigate whether these analogies
extend beyond the tree level, i.e., whether the mode-mode coupling effects in
helimagnets are as strong as in chiral liquid crystals. We find that this is
indeed the case. Some of the most interesting and dramatic effects are the
fluctuation renormalizations of the transport coefficients that determine, for
example, the line width of the neutron scattering signal due to the helical
Goldstone mode. At $T>0$ we find that fluctuation effects are so strong that
they lead to infinite renormalizations of the transport coefficients and
elastic constants. Physically, this implies that at long length scales, and for
low frequencies, the actual macroscopic description is nonlocal. In the quantum
or zero temperature limit we find that the divergencies are removed due to the
structure of the fluctuation-dissipation theorem, and the transport
coefficients and elastic constants saturate, albeit at a very low temperature.
As we discuss below, our predictions will be observable in precision neutron
scattering experiments.

Our starting point is a phenomenological hydrodynamic theory
\cite{Landau_Lifshitz_VI_1987, Ma_1976} for fluctuations in helimagnets. We
will take into account the most important nonlinear terms, and will show that
the nonlinearities qualitatively affect the physics. Let the mean magnetization
$\langle {\bm m}\rangle$ lie in the $x$-$y$ plane, with the axis of the helix
pointing in the $z$-direction:
\be
\langle{\bm m}\rangle({\bm x}) = m_0\, (\cos qz,\sin qz,0),
\label{eq:1}
\ee
with $q$ the pitch wave number of the helix. Typically, $q \ll k_{\text{F}}$,
with $k_{\text{F}}$ the Fermi wave number. The fluctuations about this average
state are described in terms of a dimensionless generalized phase variable
$u({\bm x})$, which describes the fluctuations of the $x$ and $y$-components of
${\bm m}$, and the $z$-component of the magnetization, $m_{z}({\bm x}) \equiv
m({\bm x})$ \cite{Belitz_Kirkpatrick_Rosch_2006}. The course-grained
Hamiltonian for these variables takes the form
\bea
H[u,m] &=& \frac{1}{2} \int d{\bm x}\ \left[aq^2m^2 - 2hm\right] \hskip 70pt
\nonumber\\
&& \hskip -55pt + \frac{1}{2} \int d{\bm x}\ \left[c_z\left(\partial_z u -
\frac{1}{2q}\,\left({\bm\nabla}u\right)^2\right)^2 +
c_{\perp}\left({\bm\nabla}_{\perp}^2 u\right)^2\right].
\label{eq:2}
\eea
Here $a$ is the coefficient of the square-gradient term in a purely magnetic
Hamiltonian, $h$ is the magnetic field conjugate to $m$, $c_z$ and $c_{\perp}$
are elastic constants, and ${\bm\nabla}_{\perp}^2 =
\partial_x^2 + \partial_y^2$. The first line in Eq.\ (\ref{eq:2}) is just the Gaussian part of
a Landau Hamiltonian for $m$; higher powers of $m$ are not important for our
purposes. The second line is the effective Hamiltonian for $u$, which is
constrained by symmetry considerations \cite{deGennes_Prost_1993,
Mazenko_Ramaswamy_Toner_1983}. Translational invariance implies that only
gradients of $u$ can enter, while rotational invariance requires that pure
rotations of the helix do no cost any energy, which restricts the combinations
of gradients that can enter the Hamiltonian. Up to quartic terms in $u$, Eq.\
(\ref{eq:2}) is the most general Hamiltonian consistent with these constraints.

Equation (\ref{eq:2}) completely describes the thermodynamics of the system via
the partition function $Z = \Tr \exp(-H/T)$ (we use units such that
$k_{\text{B}}=1$). The dynamics are described by coupled dynamical equations
that follow \cite{Belitz_Kirkpatrick_Rosch_2006} from the usual time-dependent
Ginzburg-Landau theory for the magnetization \cite{Ma_1976} and read
\bse
\label{eqs:3}
\bea
\partial_{\,t} u &=& \gamma\, \frac{\delta H}{\delta m} - \Gamma\,'_{\!u}\,\frac{
\delta H}{\delta u} + \zeta_u,
\label{eq:3a}\\
\partial_{\,t} m &=& -\gamma\, \frac{\delta H}{\delta u}
  + \left[\Gamma\,'_{\!z}\, \partial_z^2 + \Gamma\,'_{\!\perp}
    {\bm \nabla}_{\perp}^{\,2}\right]\,
  \frac{\delta H}{\delta m} + \zeta.
\label{eq:3b}
\eea
\ese
Here $\gamma$ is a constant, and so are the transport coefficients
$\Gamma\,'_{\!u,z,\perp}$ \cite{disorder_footnote}. $h$ is the (magnetic) field
conjugate to $m$, and $\zeta_u$ and $\zeta$ are Langevin forces. In Eq.\
(\ref{eq:3a}) the dissipative term $\propto \Gamma\,'_{\!u}$, as well as the
Langevin force $\zeta_u$, turn out to be unimportant for the long-wavelength
effects considered here, and we will neglect them in what follows.

Consider the magnetic susceptibility or linear response function $\chi$,
defined by the response of the average magnetization to an applied magnetic
field,
\be
\langle m({\bm k},\omega)\rangle = \chi({\bm k},\omega)\,h({\bm k},\omega).
\label{eq:4}
\ee
The Langevin force $\zeta$ is fixed by requiring that the correlation function
$F_{mm}({\bm k},\omega) = \langle\vert m({\bm k},\omega)\vert\rangle$
\cite{FDT_footnote} is related to the imaginary part or spectrum of $\chi$,
$\chi''({\bm k},\omega)$, by the fluctuation-dissipation theorem
\cite{Forster_1975}
\be
F_{mm}({\bm k},\omega) = \chi''({\bm k},\omega)\,\coth(\hbar\omega/2T).
\label{eq:5}
\ee
The renormalizations of the elastic coefficients and transport coefficients are
determined by examining how fluctuations modify the linear response function.
To zeroth order in the non-Gaussian terms in Eq.\ (\ref{eq:2}) the latter is
easily found, from Eqs.\ (\ref{eqs:3}), to be
\bse
\label{eqs:6}
\be
\chi_0({\bm k},\omega) = \frac{\Omega_0^2({\bm k}) - i\omega\Gamma_0({\bm k})}
{aq^2\left[-\omega^2 + \Omega_0^2({\bm k}) - i\omega\Gamma_0({\bm k})\right]}.
\label{eq:6a}
\ee
Here
\be
\Omega_0({\bm k}) = \gamma\,a^{1/2}q\,\left[c_z k_z^2 + c_{\perp}{\bm
k}_{\perp}^4\right]^{1/2}
\label{eq:6b}
\ee
is the helimagnon resonance frequency, and
\be
\Gamma_0({\bm k}) = \Gamma_z k_z^2 + \Gamma_{\perp} {\bm k}_{\perp}^2,
\label{eq:6c}
\ee
\ese
with $\Gamma_{z,\perp} = aq^2\Gamma\,'_{\!z,\perp}$, is a damping coefficient.
The spectrum of $\chi$ is proportional to the neutron scattering cross-section,
and the renormalized counterpart of $\Gamma_0$ determines the line width of the
helimagnon resonance \cite{Forster_1975}. The $\langle uu \rangle$ and $\langle
mu \rangle$ correlation functions can be related to $F_{mm}$ by Fourier
transforming Eq.\ (\ref{eq:3a}) to obtain
\be
u({\bm k},\omega) = \gamma \left[ aq^2 m({\bm k},\omega) - h\right]/(-i\omega).
\label{eq:7}
\ee

\begin{figure}[b,h]
\includegraphics[width=6cm]{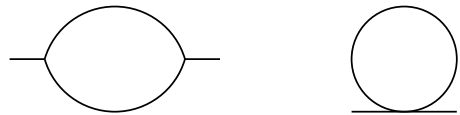}
\caption{One-loop contributions to the response function $\chi$. The two
diagrams result from the cubic and quartic in u terms, respectively, in Eq.\
(\ref{eq:2}).}
\label{fig:1}
\end{figure}

Loop corrections to $\chi_0$ due to the non-Gaussian terms in Eq.\
(\ref{eq:2}), which lead to nonlinearities in the dynamical equations, can be
calculated by means of standard perturbation theory for $\chi$. The two
one-loop diagrams are shown schematically in Fig.\ \ref{fig:1}.
It is convenient to consider the renormalization of $\chi_0^{-1}$, whose
one-loop correction we denote by $\delta \chi^{-1}$. We obtain
\bea
\delta\chi^{-1}({\bm k},\omega) &=& \frac{\omega^2}{\left[\Omega_0^2({\bm k}) -
i\omega \Gamma_0({\bm k})\right]^2}\,\left[ I_3({\bm k},\omega) + I_4({\bm
k},\omega) \right. \nonumber\\
&& \hskip 100pt \left.- C({\bm k})\right],
\label{eq:8}
\eea
where
\be
C({\bm k}) = T\Lambda^2 {\bm k}_{\perp}^2\,\frac{c_z^{1/2}(\gamma
aq^2)^2}{16\pi q^2c_{\perp}^{1/2}},
\label{eq:9}
\ee
and
\begin{widetext}
\bse
\label{eqs:10}
\bea
I_3({\bm k},\omega) &=& \frac{c_z^2(\gamma aq^2)^6}{4aq^4}\,
   \frac{1}{V}\sum_{{\bm k}_1}\ W_3({\bm k},{\bm k}_1) \int_{-\infty}^{\infty}\frac{d\omega_1}{2\pi}\
\ \frac{F_{mm}({\bm k}-{\bm k}_1,\omega -\omega_1)} {(\omega -
\omega_1)^2\,\left[-\omega_1^2 + \Omega_0^2({\bm k}_1) - i\omega_1\Gamma_0({\bm
k}_1)\right]}\ ,
\label{eq:10a}\\
I_4({\bm k},\omega) &=& \frac{c_z(\gamma aq^2)^4}{4q^2}\,
   \frac{1}{V}\sum_{{\bm k}_1}\ W_4({\bm k},{\bm k}_1)
\int_{-\infty}^{\infty}\frac{d\omega_1}{2\pi}\ \frac{F_{mm}({\bm
k}_1,\omega_1)}{\omega_1^2}\ .
\label{eq:10b}
\eea
\ese
\end{widetext}
Here
\bse
\label{eqs:11}
\bea
W_3({\bm k},{\bm p}) &=& \left[V_3({\bm k},{\bm p}) + V_3({\bm k},{\bm k}-{\bm
p})\right]\ \nonumber\\
&&\times\left[V_3({\bm p},{\bm k}) + V_3({\bm p},{\bm p}-{\bm k})\right],
\label{eq:11a}\\
W_4({\bm k},{\bm p}) &=& V_4({\bm k},{\bm k},{\bm p}) + V_4({\bm k},{\bm
p},{\bm k}) \nonumber\\
&& + V_4({\bm k},-{\bm p},{\bm p}),
\label{eq:11b}
\eea
with
\bea
V_3({\bm k},{\bm p}) &=& \left(k_z {\bm p}_{\perp} + 2p_z {\bm
k}_{\perp}\right)\cdot\left({\bm k}_{\perp} - {\bm p}_{\perp}\right),
\label{eq:11c}\\
V_4({\bm k},{\bm p}_1,{\bm p}_{\,2}) &=& \left[\left({\bm k}\cdot{\bm
p}_{1\perp}\right){\bm p}_{\,2\perp} + \left({\bm p}_{1\perp}\cdot{\bm
p}_{\,2\perp}\right){\bm k}_{\perp}\right]
\nonumber\\
&& \cdot\left({\bm p}_{1\perp} + {\bm p}_{\,2\perp} - {\bm k}_{\perp}\right).
\label{eq:11d}
\eea
\ese
The first and second terms on the right-hand side of Eq.\ (\ref{eq:8})
correspond to the two diagrams shown in Fig.\ \ref{fig:1}. The third term
arises from a counterterm that needs to be added to the Hamiltonian in order to
ensure that one expands about a proper saddle point, rather than its mean-field
approximation \cite{Grinstein_Pelcovits_1982}. $\Lambda$ is an ultraviolet
cutoff common to all momentum integrals. The technical effect of the
counterterm is to subtract the ${\bm k}_{\perp}^2$ contributions that arise in
perturbation theory and to ensure that the resonance frequency retains the
structure given in Eq.\ (\ref{eq:6b}), which is dictated by rotational
symmetry.

Using Eqs. (\ref{eq:5},\ref{eqs:6}) in Eqs.\ (\ref{eq:8}) - (\ref{eqs:11})
leads to the leading order fluctuation renormalization of the response
function. Note the difference between the classical and quantum cases: In the
classical limit, the $\coth$ in Eq. (\ref{eq:5}) contributes a factor
$T/(\omega -\omega_1)$ to the integrand in Eq.\ (\ref{eq:10a}), while in the
quantum limit it contributes a step function. The important implication is that
in the quantum limit, the analog of classical mode coupling effects are
relatively less singular at low frequencies.

Equation (\ref{eq:8}) effectively defines a renormalization of the elastic
coefficients and the transport coefficients \cite{m4_footnote}. Denoting the
corrections to $\Omega_0^2$ and $\Gamma_0$ by $\delta\Omega^2$ and
$\delta\Gamma$, respectively, we find
\bse
\label{eqs:12}
\bea
\delta\Omega^2({\bm k}) &=& \frac{1}{aq^2}\ \text{Re}\left[ I_3({\bm k},0) +
I_4({\bm k},0) - C({\bm k})\right],\hskip 20pt
\label{eq:12a}\\
\delta\Gamma({\bm k},\omega) &=& \frac{-1}{aq^2\omega}\ \text{Im}\, I_3({\bm
k},\omega).
\label{eq:12b}
\eea
\ese

Equations (\ref{eqs:12}) are the central results of this paper. In the
classical limit, $\hbar=0$, at zero external wave number and finite frequency
the perturbative corrections to the transport coefficients for small
$\Gamma_{\perp}$ are
\bse
\label{eqs:13}
\bea
\delta \Gamma_z &=& \frac{(\gamma aq^2)^2 c_z^{3/2}}{128 aq^4
c_{\perp}^{3/2}}\,\frac{T}{\vert\omega\vert}\ ,
\label{eq:13a}\\
\delta \Gamma_{\perp} &=& \frac{(\gamma aq^2)^2 c_z^{1/2}}{32\pi aq^4
c_{\perp}^{1/2} \Gamma_{\perp}}\,T\,\ln
(\Lambda^2\Gamma_{\perp}/\vert\omega\vert)\ .
\label{eq:13b}
\eea
\ese
For the one-loop corrections to the elastic coefficients at zero frequency we
find
\bse
\label{eqs:14}
\bea
\delta c_{z} &=& -\frac{c_{z}^{3/2}T}{32\pi q^{2}c_{\perp}^{3/2}}\ \ln
(\Lambda^2/k_{\perp}^2)\ ,
\label{eq:14a}\\
\delta c_{\perp} &=& \frac{c_{z}^{1/2}T}{64\pi q^{2}c_{\perp}^{1/2}}\ \ln
(\Lambda^2/k_{\perp}^2)\ .
\label{eq:14b}
\eea
\ese
Equations (\ref{eqs:14}) can also be obtained from equilibrium statistical
mechanics considerations, by renormalizing the Hamiltonian, Eq.\ (\ref{eq:2}).
Note that $\Gamma_z,$ $\Gamma_{\perp},$ and $c_{\perp}$ are renormalized to
infinity, while $c_z$ is renormalized to zero \cite{resummation_footnote}. Also
note that the fluctuation corrections are all linearly dependent on the
temperature. This is due to the fact that all of the excitations are bosonic in
character.

Physically, the Eqs. (\ref{eqs:14}) imply that the helimagnon state is,
strictly speaking, not stable at finite temperatures. The same is true for
certain liquid crystals that show one-dimensional order, and it shows that the
analogies between helimagnets and liquid crystals pointed out in Ref.\
\cite{Belitz_Kirkpatrick_Rosch_2006} at a Gaussian level do indeed carry over
to the respective nonlinear theories. This is a nontrivial observation, since
the dynamical equations are different in the two cases. However, practically
speaking one must go to enormous length or time scales for these divergencies
to dominate the zero-loop effects. In liquid crystals the singularity is
typically cut off by finite system size effects; in a helimagnet, it will be
cut off by crystal-field effects that break the rotational symmetry. The
strongest effect is the renormalization of $\Gamma_z$, which shows a linear
divergency rather than a logarithmic one.

We now turn to the quantum or zero-temperature limit, where the $\coth$ in Eq.\
(\ref{eq:5}) turns in to a step function. $\delta\Gamma_{\perp}$, $\delta c_z$,
and $\delta c_{\perp}$ are then finite by power counting. $\delta\Gamma_z$ one
might expect to still be logarithmically divergent, but a detailed inspection
of the relevant integral shows that it, too, is finite. We find
\be
\delta\Gamma_z = \frac{(\gamma aq^2)^2 c_z^{3/2}}{512\,aq^4
c_{\perp}^{3/2}}\quad,\quad (T=0)\ .
\label{eq:15}
\ee
Comparing with Eq.\ (\ref{eq:14a}) we see that $\delta\Gamma_z$ saturates at a
value corresponding to $T=\vert\omega\vert/4$. All of the classical
divergencies thus vanish at $T=0$.

Let us finally provide order-of-magnitude estimates for the magnitude of these
fluctuation effects. The coefficients $a$ (which dimensionally is a length
squared) and $\gamma^2$ (which dimensionally is an energy times a length cubed)
have nothing fundamental to do with magnetism and are of order $a \sim
1/k_{\text{F}}^2$ and $\gamma^2 \sim \epsilon_{\text{F}}/k_{\text{F}}^3$,
respectively. We denote the Fermi energy, wave number, and velocity by
$\epsilon_{\text{F}}$, $k_{\text{F}}$, and $v_{\text{F}}$, respectively, and
the Stoner gap or exchange splitting by $\lambda$. The tilde indicates that we
are leaving out a numerical prefactor that one expects to be close to unity.
For the elastic coefficients one expects $c_z \sim c_{\perp} q^2 \propto
\lambda^2/v_{\text{F}}$. However, the model calculation of Ref.\
\onlinecite{Belitz_Kirkpatrick_Rosch_2006} yielded a small prefactor on the
order of $10^{-3}$ for these coefficients, and we will use $c_z = 2 c_{\perp}
q^2 \sim 0.001\, \lambda^2/v_{\text{F}}$ for our estimates. The transport
coefficients $\Gamma\,'{\!_z}$ and $\Gamma\,'_{\!\perp}$ have again nothing to
do with magnetism, and we expect $\Gamma\,'_{\!z} \sim \Gamma\,'_{\!\perp} \sim
v_{\text{F}}^2\tau$, with $\tau$ the electronic elastic mean-free time. This
leads to $\Gamma_z \sim \Gamma_{\perp} \sim (q/k_{\text{F}})^2
v_{\text{F}}^2\tau$.

Now consider the singularities at $T>0$. In a solid-state system they are cut
off at a length scale $L$ that is inversely proportional to the spin-orbit
coupling squared, or, since $q$ is proportional to the spin-orbit coupling, $L
\sim 2\pi k_{\text{F}}/q^2$ \cite{Belitz_Kirkpatrick_Rosch_2006}. The cutoff
frequency for the dynamical singularities is given by $\Omega_0(k\approx 1/L)$.
With these estimates we obtain
\bse
\label{eqs:16}
\bea
\frac{\delta\Gamma_z}{\Gamma_z} &\sim& \frac{T}{\epsilon_{\text{F}}}\,
\frac{\epsilon_{\text{F}}}{\lambda}\,\frac{L}{\ell}\ ,
\label{eq:16a}\\
\frac{\delta\Gamma_{\perp}}{\Gamma_{\perp}} &\sim& 0.003\,
\frac{T}{\epsilon_{\text{F}}}\,\left(\frac{k_{\text{F}}}{q}\right)^3\,
\frac{1}{(k_{\text{F}}\ell)^2}\,\ln(k_{\text{F}}L),
\label{eq:16b}
\eea
\ese
with $\ell = v_{\text{F}}\tau$ the elastic mean-free path. For the elastic
constants we find
\be
\frac{\delta c_z}{c_z} \sim - \frac{\delta c_{\perp}}{c_{\perp}} \sim - 100\,
\frac{T}{\epsilon_{\text{F}}}\,\left(\frac{\epsilon_{\text{F}}}{\lambda}\right)^2\,
\frac{q}{k_{\text{F}}}\,\ln(k_{\text{F}}L),
\label{eq:17}
\ee
Parameter values for MnSi are (\cite{Belitz_Kirkpatrick_Rosch_2006} and
references therein) $\epsilon_{\text{F}} \approx 23,000\,{\text{K}}$,
$q/k_{\text{F}} \approx 0.024$, and $L \approx 7,500\,\AA$. The value of
$\epsilon_{\text{F}}/\lambda$ is less certain, we will assume
$\epsilon_{\text{F}}/\lambda = 4$, which leads to $\delta c_z/c_z \approx 0.1$
at $T=10\,{\text{K}}$. The corrections to the damping coefficients are much
smaller. With a mean-free path $\ell \approx 5,000\,\AA$
\cite{Pfleiderer_Julian_Lonzarich_2001} one finds $\delta\Gamma_z/\Gamma_z
\approx 10^5 \delta\Gamma_{\perp}/\Gamma_{\perp} \approx 10^{-3}$ at $T =
10\,{\text{K}}$. In less clean samples the fluctuations of $\Gamma_z$ and
$\Gamma_{\perp}$ will obviously be larger.

The temperature scale where the divergent classical results cross over to the
finite expressions in the quantum limit follows from Eqs.\
(\ref{eq:13a},\ref{eq:15}), and is corroborated by an inspection of the
relevant integrals. It is given by
\be
T^* = \Omega_0(k_z = k_{\perp} = 1/L) \approx \gamma a^{1/2}q c_z^{1/2}/L.
\label{eq:18}
\ee
With the same parameter values as above we find $T^* \approx 0.3\,{\text{mK}}$.
In interpreting all of these quantitative results one needs to keep in mind
that these are {\em very} rough estimates that depend strongly on some of the
parameters.

In conclusion, we find anomalously large fluctuation effects in the ordered
phase of helimagnets which lead to a breakdown of hydrodynamics. In MnSi, the
largest observable signature is predicted to be a linear-in-$T$ shift in the
resonance frequency squared, with an absolute value of the shift of about 10\%
at $T=10\,{\text{K}}$. This should be observable by neutron scattering.

This work was supported by the NSF under grant Nos. DMR-05-29966 and
DMR-05-30314. We are grateful to John Toner for useful discussions.

\vskip -1mm

\begin{thebibliography}{17}
\bibitem[{dis()}]{disorder_footnote}
\bibinfo{note}{In a perfectly clean system at zero temperature the transport
  coefficients are functions of frequency and wave number that diverge in the
  limit of zero frequency and wave number. We assume a small amount of quenched
  disorder to be present that makes these coefficients finite without having
  any other effects.}
\bibitem[{FDT()}]{FDT_footnote}
\bibinfo{note}{Notice that, in the framework of flucuating hydrodynamics,
  $m({\bm k},\omega)$ is c-number valued, and the relevant correlation function
  is therefore the fluctuation function or symmetrized correlation function
  $F$, which is an even function of the frequency, rather than the van Hove
  function.}
\bibitem[{m4_()}]{m4_footnote}
\bibinfo{note}{$aq^2$, which is the coefficient of $m^2$ in Eq.\ (\ref{eq:2}),
  is not renormalized since we have omitted terms non-Gaussian in $m$ from the
  action.}
\end{thebibliography}

\end{document}